\documentstyle[aps]{revtex}

\draft

\begin{document}

\tighten

\twocolumn[\hsize\textwidth\columnwidth\hsize\csname@twocolumnfalse\endcsname

\title{Marginal Fermi liquid resonance induced by a quantum magnetic impurity in
d-wave superconductors}
\author{Guang-Ming Zhang$^{1,2}$, Hui Hu$^2$, Lu Yu$^{3,4}$}
\address{$^1$Center for Advanced Study, Tsinghua University, Beijing 100084, China\\
$^2$Department of Physics, Tsinghua University, Beijing 100084, China\\
$^3$Abdus Salam International Center for Theoretical Physics, P. O. Box 586, Trieste
34100, Italy\\
$^4$Institute of Theoretical Physics, Academic Sinica, Beijing 100080, China}
\date{\today}
\maketitle

\begin{abstract}
We consider a model of an Anderson impurity embedded in a $d_{x^2-y^2}$-wave
superconducting state to describe the  low-energy 
excitations of  cuprate superconductors  doped with a small amount
of magnetic impurities. Due to the Dirac-like energy dispersion,
a sharp localized resonance above the Fermi energy, showing   a
marginal Fermi liquid behavior ($\omega \ln \omega$ as $\omega \rightarrow 0$),
 is predicted for the impurity states. 
The same logarithmic dependence of self-energy  and 
a linear frequency dependence of the relaxation rate are also derived
for the conduction electrons, characterizing a new universality class
for the strong coupling fixed point.  At the resonant energies, the spatial
distribution of the electron density of states around the magnetic impurity is
calculated, to be confronted with  measurements of  the scanning
tunneling microscopy on Bi$_2$Sr$_2$Ca(Cu$_{1-x}$Ni$_x$)O$_{8+\delta }$.
\end{abstract}


]

It has been for a long time that non-magnetic and magnetic impurities are
exploited to elucidate the microscopic nature of the superconducting state.
In the high-T$_c$ cuprates, substitution by divalent metals (Zn and Ni) for
Cu in the CuO$_2$ plane offers a particularly important way of introducing
such impurities, as they preserve the doping level and introduce only
minimal structural disorder. Recently, scanning tunneling microscopy (STM)
has been further developed to probe the quasiparticle scattering states
around a single Zn impurity in Bi$_2$Sr$_2$Ca(Cu$_{1-x}$Zn$_x$)O$_{8+\delta
} $ with a high spatial and energy resolution by Pan {\it et al}.\cite
{pan-nature}. In the obtained STM spectra, an intense zero-bias
quasiparticle scattering resonance is found at the Zn sites, and the spatial
dependence of the density of states (DOS) in the vicinity of the impurities
reveals a fourfold symmetry which characterizes $d_{x^2-y^2}$ wave
superconductivity (dSC) \cite{pan-nature}. In fact, the existence of a
non-magnetic impurity induced resonant state in dSC was predicted
theoretically by Balatsky {\it et al}.\cite{balatsky,salkola,sbs} earlier.

Here, we consider the scattering effects of magnetic impurities on dSC. In
Ni-doped cuprates, in order to maintain the valence of Cu$^{2+}$ ions (3d$^9$%
, S=1/2), the Ni$^{2+}$ ions have a configuration (3d$^8$, S=1), and strong
antiferromagnetic exchange couplings with the neighboring Cu sites lead to a
residual S=1/2 on the Ni site, acting as a localized magnetic spin weakly
coupled to its environment through exchange interactions \cite{sidis}.
Although there are some theoretical studies on the quasiparticle states
around such a magnetic impurity in dSC state \cite{sbs,flatte,hiroki,fradkin}%
, most of them treated a {\it classical} magnetic impurity. It is thus
interesting and timing to study how a {\it quantum} magnetic impurity
affects the quasiparticle states in optimally doped Bi$_2$Sr$_2$Ca(Cu$_{1-x}$%
Ni$_x$)O$_{8+\delta }$ superconductors. Since the quantum fluctuations of
the internal degrees of freedom of the magnetic impurity play an important
role in the ordinary quantum impurity problems, one can thus expect that the
effects of a {\it quantum} magnetic impurity are significantly different
from those of a {\it classical} magnetic impurity. Within the slave boson
mean field (MF) theory, we predict a sharp resonance {\it above} the Fermi
level and a marginal Fermi liquid (MFL) behavior for both impurity and
surrounding conduction electrons. Moreover, we explicitly calculate the
spatial distribution of conduction electron DOS to be compared with STM
measurements.

In this Letter, we assume that a BCS-type weak coupling theory is applicable
as a phenomenological model for High-T$_c$ optimally doped superconductors
though the underlying mechanisms are different. We also assume the magnetic
impurities are described by the Anderson model with a strong Hubbard
repulsion. When the correlations between the magnetic impurities on
different sites are ignored, the model Hamiltonian is given by

\begin{eqnarray}
{\cal H} &=&\sum\limits_{{\bf k}\sigma }\epsilon _{{\bf k}}c_{{\bf k}\sigma
}^{+}c_{{\bf k}\sigma }+\sum\limits_{{\bf k}}\Delta _{{\bf k}}\left( c_{{\bf %
k}\uparrow }^{+}c_{-{\bf k}\downarrow }^{+}+h.c.\right) +\epsilon
_d\sum\limits_\sigma d_\sigma ^{+}d_\sigma  \nonumber \\
&&+V\sum\limits_{{\bf k}\sigma }\left( c_{{\bf k}\sigma }^{+}d_\sigma
+h.c.\right) +Ud_{\uparrow }^{+}d_{\uparrow }d_{\downarrow
}^{+}d_{\downarrow }.
\end{eqnarray}
where $\epsilon _{{\bf k}}=\hbar ^2k^2/(2m)-\epsilon _F$ is the dispersion
of the conduction electrons, $\Delta _{{\bf k}}=\Delta _0\cos 2\varphi $ is
the dSC order parameter, and $\Delta _0$ is the gap amplitude. When Nambu
spinors are introduced 
\[
\hat{\psi}_{{\bf k}}=\left( 
\begin{array}{c}
c_{{\bf k}\uparrow } \\ 
c_{-{\bf k}\downarrow }^{+}
\end{array}
\right) ,\qquad \widehat{\varphi }=\left( 
\begin{array}{c}
d_{\uparrow } \\ 
d_{\downarrow }^{+}
\end{array}
\right) , 
\]
we can simplify the model Hamiltonian in a matrix form 
\begin{eqnarray}
{\cal H} &=&\sum\limits_{{\bf k}}\hat{\psi}_{{\bf k}}^{+}\left( \epsilon _{%
{\bf k}}\sigma _z+\Delta _{{\bf k}}\sigma _x\right) \hat{\psi}_{{\bf k}%
}+V\sum\limits_{{\bf k}}\left( \hat{\psi}_{{\bf k}}^{+}\sigma _z\widehat{%
\varphi }+h.c.\right)  \nonumber \\
&&+(\epsilon _d+\frac U2)(\widehat{\varphi }^{+}\sigma _z\widehat{\varphi }%
+1)-\frac U2(\widehat{\varphi }^{+}\widehat{\varphi }-1)^2,
\end{eqnarray}
where $\sigma _z$ and $\sigma _x$ are Pauli matrices. Using the method of
equations of motion, the generalized {\bf T}-matrix is derived $\hat{T}%
\left( \omega +i0^{+}\right) =V\sigma _z\hat{G}_d\left( \omega
+i0^{+}\right) V\sigma _z$, where $\hat{G}_{{\bf k}}^0\left( i\omega
_n\right) =\left( i\omega _n-\epsilon _{{\bf k}}\sigma _z-\Delta _{{\bf k}%
}\sigma _x\right) ^{-1}$ is the unperturbed Green function (GF) of the
conduction electrons. At zero temperature, analytical continuation is used
to calculate the perturbed GF through the GF of the impurity: $\hat{G}\left( 
{\bf r},{\bf r}^{\prime };\omega \right) =\hat{G}^0\left( {\bf r}-{\bf r}%
^{\prime },\omega \right) +\hat{G}^0\left( {\bf r},\omega \right) \hat{T}%
\left( \omega \right) \hat{G}^0\left( -{\bf r}^{\prime },\omega \right) $.
The local DOS of the conduction electrons around the magnetic impurity is
thus given by $N\left( {\bf r},\omega \right) =-\frac 1\pi 
\mathop{\rm Im}%
\hat{G}_{11}\left( {\bf r},{\bf r};\omega \right) $, and the relaxation rate
for the conduction electrons is also deduced from {\rm Im}$T_{11}(\omega
+i0^{+})$.

When we take the infinite $U$ limit, the impurity operator is expressed as $%
\widehat{\varphi }^{+}=\left( 
\begin{array}{cc}
f_{\uparrow }^{+}b, & f_{\downarrow }b^{+}
\end{array}
\right) $ in the slave-boson representation \cite{barnes,coleman}, where the
fermion $f_\sigma $ and the boson $b$ describe the singly occupied and hole
states, respectively. The constraint $b^{+}b+\sum\limits_\sigma f_\sigma
^{+}f_\sigma =1$ has to be imposed. When a MF approximation is applied, the
boson operators $b$ and $b^{+}$ are replaced by a c-number $b_0$, and the
constraint is satisfied by introducing a chemical potential $\lambda _0$.
Therefore, the MF Hamiltonian is written as 
\begin{eqnarray}  \label{hmf}
{\cal H}_{mf} &=&\sum\limits_{{\bf k}}\hat{\psi}_{{\bf k}}^{+}\left(
\epsilon _{{\bf k}}\sigma _z+\Delta _{{\bf k}}\sigma _x\right) \hat{\psi}_{%
{\bf k}}+\tilde{V}\sum\limits_{{\bf k}}\left( \hat{\psi}_{{\bf k}}^{+}\sigma
_z\widehat{\phi }+h.c.\right)  \nonumber  \label{Hmf} \\
&&+\tilde{\epsilon}_d\widehat{\phi }^{+}\sigma _z\widehat{\phi }+\tilde{%
\epsilon}_d+\lambda _0(b_0^2-1).
\end{eqnarray}
where $\widehat{\phi }^{+}=\left( 
\begin{array}{cc}
f_{\uparrow }^{+}, & f_{\downarrow }
\end{array}
\right) $denotes the Nambu spinors of the impurity quasiparticles and the
renormalized parameters $\tilde{\epsilon}_d=\epsilon _d+\lambda _0$ and $%
\tilde{V}=b_0V$.

Using the standard techniques we find $\hat{G}_f(i\omega _n)=\left[ i\omega
_n-\tilde{\epsilon}_d\sigma _z-\hat{\Sigma}_f\left( i\omega _n\right)
\right] ^{-1}$, where the self-energy of the impurity becomes diagonal, 
\begin{equation}
\Sigma _f\left( i\omega _n\right) =-i\omega _n\sum\limits_{{\bf k}}\frac{%
\tilde{V}^2}{\omega _n^2+\epsilon _{{\bf k}}^2+\Delta _{{\bf k}}^2},
\end{equation}
because of the inversion symmetry in the {\bf k} summation.

At $T=0$, the ground-state energy change due to impurity is:

\[
{\cal E}_{imp}=\tilde{\epsilon}_d+\lambda _0\left( b_0^2-1\right) -\frac 1\pi
\int\limits_0^Wd\omega \ln \left[ \omega ^2\left( 1+\alpha \left( \omega
\right) \right) ^2+\tilde{\epsilon}_d^2\right] , 
\]
where $W$ is the band width and $\alpha \left( \omega \right) =\sum\limits_{%
{\bf k}}\frac{\tilde{V}^2}{\omega ^2+\epsilon _{{\bf k}}^2+\Delta _{{\bf k}%
}^2}$. The saddle-point equations are derived as 
\begin{eqnarray}
\lambda _0 &=&\frac 1\pi \int\limits_0^Wd\omega \frac{2\omega ^2\left(
1+\alpha \left( \omega \right) \right) \alpha \left( \omega \right) }{\left[
\omega ^2\left( 1+\alpha \left( \omega \right) \right) ^2+\tilde{\epsilon}%
_d^2\right] b_0^2}, \\
b_0^2 &=&\frac 1\pi \int\limits_0^Wd\omega \frac{2\tilde{\epsilon}_d}{\left[
\omega ^2\left( 1+\alpha \left( \omega \right) \right) ^2+\tilde{\epsilon}%
_d^2\right] }.
\end{eqnarray}
Solving these equations yields $b_0$ and $\lambda _0$ for given parameters $%
W $, $\Delta _0$, $\Gamma =\pi N_FV^2$, and $\epsilon _d$, where $N_F$ is
the DOS at the Fermi surface. In the following, we will choose $\Delta _0$
as the energy unit, $W/\Delta _0=20$, and $\Gamma /\Delta _0=0.2$.

In the present model the local DOS of quasiparticles near the Fermi surface
goes to zero {\it linearly}, so the usual logarithmic Kondo singularity in
the scattering matrix of the magnetic moment with conduction electrons is
thus absent. When $\epsilon _d$ is less than a threshold value, $b_0^2$ is
zero, leading to a decoupled free local magnetic moment, namely, no Kondo
effect occurs. However, above the threshold value of $\epsilon _d$, $b_0^2$
rises steeply, and then saturates quickly. The usual broad mixed valence
regime shrinks to a very narrow regime. In Fig. 1, the ground state phase
diagram is calculated in the $\epsilon _d-\Gamma $ plane. For a given value
of $\epsilon _d$, there will be a phase transition from the decoupled free
spin to the mixed valence, and finally a crossover to the strong coupling
regime. The finite threshold value of the phase transition is delineated by
the solid line between area {\rm I} and {\rm II,} and turns out to be linear
in $\left| \epsilon _d\right| /\Delta _0$, approximately.

In the mixed valence and strong coupling regimes, the impurity DOS versus $%
\epsilon _d$ is plotted in Fig. 2. A sharp local resonance always appears 
{\it above} the Fermi energy for each value of $\epsilon _d$, while the
corresponding DOS for $\omega <\;0$ is broad and small. This is one of the
most important differences between the magnetic and non-magnetic impurities
scatterings, as the localized resonance always occurs {\it below} the Fermi
energy for the repulsive potential scattering in the latter case \cite
{balatsky}. To the logarithmic accuracy, the zero of the denominator of $%
\hat{G}_f(i\omega _n)$ is given by $\Omega =\Omega ^{\prime }-i\Omega
^{\prime \prime }$, and 
\[
\Omega ^{\prime }\approx \tilde{\epsilon}_d\left[ 1-b_0^2\left( \frac{%
2\Gamma }{\pi \Delta _0}\right) \ln \frac{4\Delta _0}{\tilde{\epsilon}_d}%
\right] ,\text{ \quad }\Omega ^{\prime \prime }\approx \Omega ^{\prime
}b_0^2\left( \frac \Gamma {\Delta _0}\right) , 
\]
where $\Omega ^{\prime }$ represents the position of the quasiparticle
resonance, while $\Omega ^{\prime \prime }$ corresponds to its width or the
inverse lifetime. If the self-energy $\Sigma _f\left( \omega \right) $ is
expanded near the resonant energy, the impurity DOS can be approximately
written in a Lorentzian form $N_{imp}\left( \omega \right) \approx \frac 1\pi
\frac{b_0^2\Omega ^{\prime \prime }}{\left( \omega -\Omega ^{\prime }\right)
^2+\left( \Omega ^{\prime \prime }\right) ^2}$. At the resonant energy $%
\omega =\Omega ^{\prime }$ the height of the resonance is $\frac{\Delta _0}{%
\pi \Gamma }\frac 1{\Omega ^{\prime }}$, inversely proportional to the
resonant energy. As $\tilde{\epsilon}_d\rightarrow 0$, this resonance
becomes arbitrarily sharp and close to the Fermi surface, but the DOS at the
Fermi energy is {\it always} suppressed to zero for all values of $\epsilon
_d$ because of the imaginary part of the impurity self-energy.

Actually, an analytic expression for the retarded self-energy of the
magnetic impurity can be derived, 
\[
\Sigma _f\left( \omega \right) =-b_0^2\left( \frac{2\Gamma }{\pi \Delta _0}%
\right) \omega \left[ K\left( \sqrt{1-\epsilon ^2}\right) +i{\rm sgn}(\omega
)K\left( \epsilon \right) \right] , 
\]
for $\epsilon \equiv \left| \omega \right| /\Delta _0<1.$ Here, $K$\ is the
complete elliptic integral of the first kind. As $\omega \rightarrow 0,$ we
have 
\begin{eqnarray}
\mathop{\rm Re}%
\Sigma _f\left( \omega \right) &\sim &-b_0^2\left( \frac{2\Gamma }{\pi
\Delta _0}\right) \omega \ln \frac{4\Delta _0}{\left| \omega \right| }, \\
\mathop{\rm Im}%
\Sigma _f\left( \omega \right) &\sim &-b_0^2\frac \Gamma {\Delta _0}\left(
\left| \omega \right| +\frac 14\frac{\left| \omega \right| ^3}{\Delta _0^2}%
\right) ,
\end{eqnarray}
{\it i.e}. precisely the MFL behavior proposed by Varma {\it et al.} to
describe the anomalous normal state properties of optimally doped cuprates 
\cite{varma}. Within the {\bf T}- matrix approximation, the self-energy of
the conduction electron has exactly the same type of singular behavior. To
our knowledge, it is the first time to obtain such a result.

Earlier, the Kondo effect in ``gapless'' fermion systems (with DOS $\rho
(\epsilon )\sim |\rho |^r,0<r\leq 1$) has been studied by a number of authors%
\cite{fradkin1}. They found a critical value for the Kondo coupling constant
below which the local moment decouples. This feature has been reconfirmed in
our calculations. However, beyond the critical value they found the same
Fermi liquid strong coupling fixed point as in the standard Kondo problem.
To the contrary, some {\it dramatically different } results are obtained in
our studies. We find a MFL behavior in the mixed valence and nearly empty
orbital regimes. Namely, the real part of the self-energy goes like $\omega
\ln \omega $, while the imaginary part behaves like $|\omega |$ as $\omega
\rightarrow 0$. We believe {\it a new universality class} for the strong
coupling fixed point has been found. The discrepancy with earlier treatments
is due to the fact that {\it different limits} are considered. In their case
the occupation of the impurity is always one and there is a true localized
energy level well below the Fermi energy. We, in contrast, are considering
the opposite limit, when the hybridization is assumed to be large and the
impurity energy levels can merge with the conduction electrons. From the
theoretical point of view, the appearance of MFL behavior in our model is
fully understandable. It is well-known that near the nodes of a dSC, a Dirac
type spectrum appears and the standard dimensional analysis of the quantum
field theory can be applied \cite{sachdev}. The scaling dimensions of the
Nambu spinors $\widehat{\Psi }({\bf r},t)$ and $\widehat{\phi }(t)$ turn out
to be $-1$ and $0$ in length units, respectively. Thus short-range
interactions between the conduction electrons are irrelevant, while the
hybridization term of the Anderson Hamiltonian is {\it marginal} and is
responsible for this MFL behavior. It seems to us that the Dirac structure
of the energy dispersion itself is the main reason behind the MFL for the
strong coupling fixed point.

Focus now on $N\left( {\bf r},\omega \right) $ in the spatial range $0<{\bf %
r\leq }$ $\xi $. Here $\xi =\hbar \upsilon _F/\Delta _0$ is the coherence
length of dSC state, and also the natural length unit of our model, while in
high T$_c$ dSC state, $\xi $ is about $10{\rm \AA }$, or roughly 3 lattice
spacings. In Fig. 3, the local DOS vs frequency is shown for $r{\bf =}%
0.07\xi $ from the magnetic impurity along the directions of the gap maxima
and the gap nodes. In addition to usual V-shape structure, there are
quasiparticle resonances near the Fermi energy, and the positions of these
resonant peaks coincide with those of the impurity resonances $\omega =\pm
\Omega ^{\prime }$. Along the directions of the gap maxima, there are two
resonances below and above the Fermi energy, which are slightly asymmetric
in the line shape. On the other hand, along the directions of the gap nodes,
there is only one sharp resonance and the local DOS is entirely hole-like.
As the impurity energy level $\epsilon _d$ increases, the quasiparticle
resonances become broader, exhibiting a similar dependence as the local
resonance of the impurity \cite{flatte}.

We also calculate the spatial variation of the DOS of the conduction
electrons. The DOS around the magnetic impurity at the resonance energies is
displayed in Fig.4a\ for $+\Omega ^{\prime }$ and in Fig.4b for $-\Omega
^{\prime }$ as a function of spatial variables for $\epsilon _d/\Delta
_0=-0.2$ in a logarithmic intensity scale. The quasiparticle resonances
induced by the magnetic impurity are highly localized around the impurity,
and the spatial oscillation of these resonant states is visible. The largest
amplitude of the quasiparticle resonance occurs at the neighborhood of the
impurity, and the local electronic structures distinctly differ in Fig.4a
and Fig.4b. For $\omega =+\Omega ^{\prime }$, the local DOS exhibits a
four-fold symmetry along the directions of the gap nodes for all distances,
consistent with the dSC of the conduction electrons. For $\omega =-\Omega
^{\prime }$, the local DOS is strongly enhanced in the gap maxima directions
at distances $r\ll \xi $. Further away from the impurity ($r{\bf \sim }\xi $%
), it is confined to the neighborhood of the diagonal directions, leading to
an eight-fold symmetry.

The logarithmic correction to the real part of the self-energy is a very
subtle effect to detect experimentally. However, its imaginary part, the
inverse quasiparticle lifetime $\tau ^{-1}\propto -(\widetilde{V}/\widetilde{%
\epsilon }_d)^2$ Im $\Sigma _f(\omega )\propto \left| \omega \right| $ can
be checked by experiments directly. Amazingly, such a linear frequency
dependence of the inverse quasiparticle lifetime has been observed in the
recent angle-resolved photoemission experiments in optimally doped Bi$_2$Sr$%
_2$CaCu$_2$O$_{8+\delta }$ along the nodal directions \cite{valla,kaminsky}.
Although these high T$_c$ cuprates are believed to contain a small number of
intrinsic defects or implications and ''impurity scatterings'' may lead to
localization of quasiparticle states, it is not clear whether the Anderson
impurity model embedded in dSC state is applicable in this case.

To conclude, we have investigated the {\it quantum} magnetic impurity
effects in high-T$_c$ superconductors based on the Anderson model. We have
found a new universality class for the strong coupling fixed point for this
type of models. We have made explicit predictions on the resonance states
around the magnetic impurities to be compared with experiments.

After submitting the manuscript, we received a preprint on optimally doped Bi%
$_2$Sr$_2$Ca(Cu$_{1-x}$Ni$_x$)O$_{8+\delta }$ from Dr. S. H. Pan \cite
{hudson} in which a localized resonance {\it above} the Fermi energy has
been reported in the DOS of the Ni impurity, and the spatial dependence of
the conduction electron DOS at the resonant energies is in a reasonable
agreement with our calculations.

One of the authors (G.-M. Zhang) would like to thank D. H. Lee, S. H. Pan,
and M. Shaw for their stimulating discussions, and also to acknowledge the
financial support from NSF-China (Grant No. 10074036) and the Special Fund
for Major State Basic Research Projects of China (G2000067107).

\begin{center}
{\bf Figures Captions}
\end{center}

Fig. 1. The ground-state phase diagram of the model. Three areas denoted by 
{\rm I, II, }and{\rm \ III }correspond to decoupled local magnetic moment,
mixed valence, and strong coupling regimes, respectively.

Fig. 2. The DOS $N_{imp}\left( \omega \right) $ of the magnetic impurity,
(a) for $\omega <0$ and (b) for $\omega >0$, with $\epsilon _d/\Delta _0=$ $%
-0.2$, $0.0$, $0.2$, and $0.4$, denoted by solid, dashed, dotted, and
dash-dotted lines, respectively.

Fig. 3. The local DOS $N\left( r,\omega \right) $ of the conduction
electrons for $\epsilon _d/\Delta _0=-0.2$, $0.0$, and $0.2$ (from top to
bottom) in units of $N_F$. Here $r{\bf =}0.07\xi $ correspond to the largest
amplitude of the quasiparticle resonance at the neighborhood of the
impurity. (a)\ along the directions of the gap maxima and (b) along the
directions of the gap nodes.

Fig. 4. The spatial distributions of the conduction electron DOS around the
impurity at the resonant energies, (a)\ $\omega =\Omega ^{\prime }$ and (b) $%
\omega =$ $-\Omega ^{\prime }$. Here $\epsilon _d/\Delta _0=-0.2$ and a
logarithmic intensity scale is used. The coherent length $\zeta $ is about $%
10{\rm \AA }$, or roughly 3 lattice spacings in high T$_c$ cuprates.

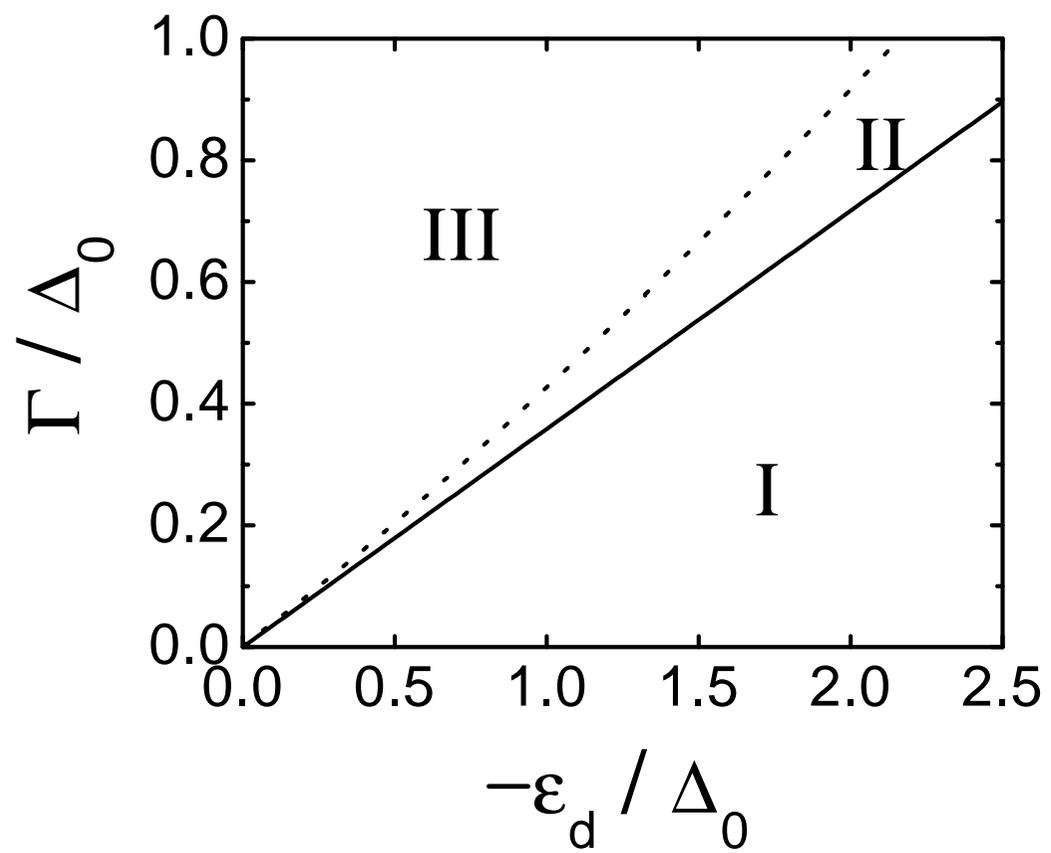
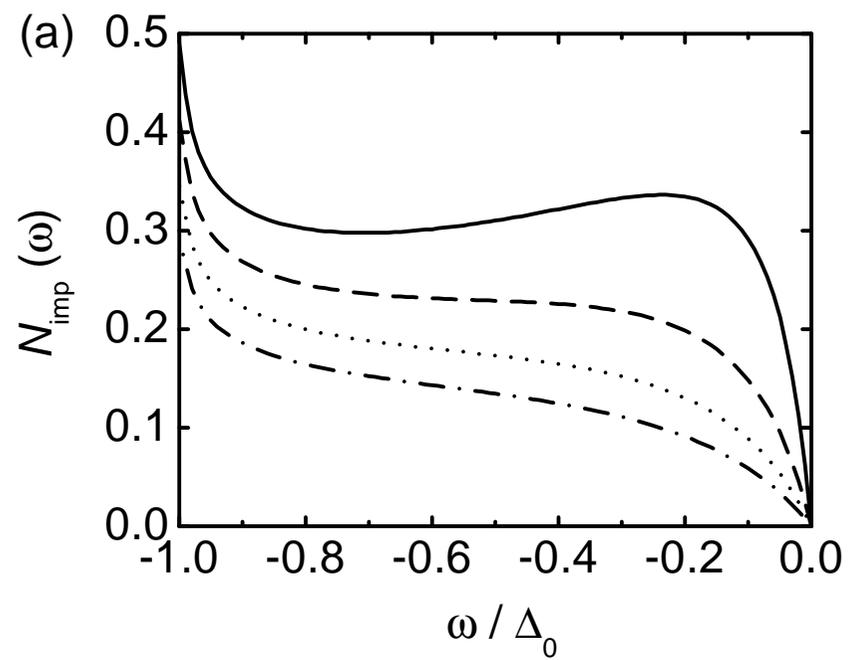
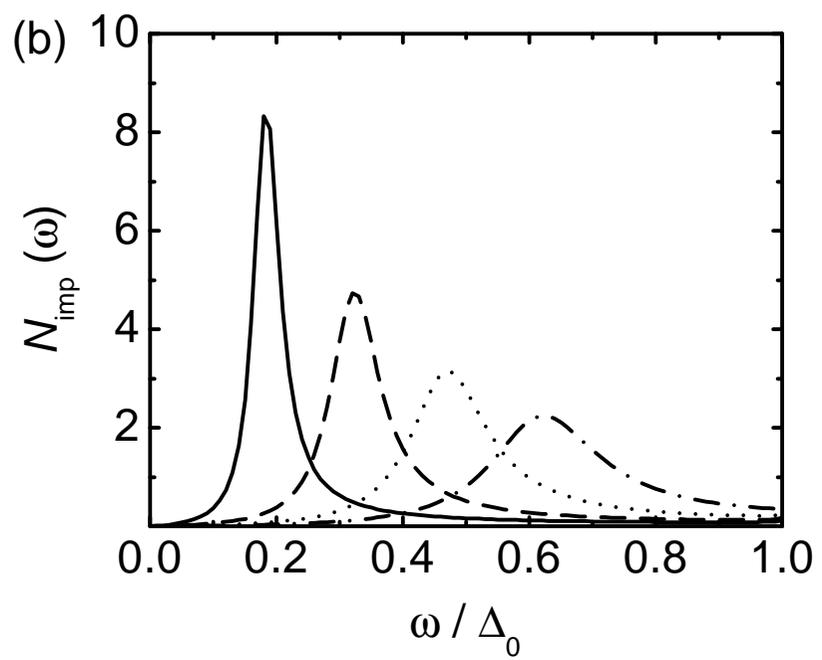
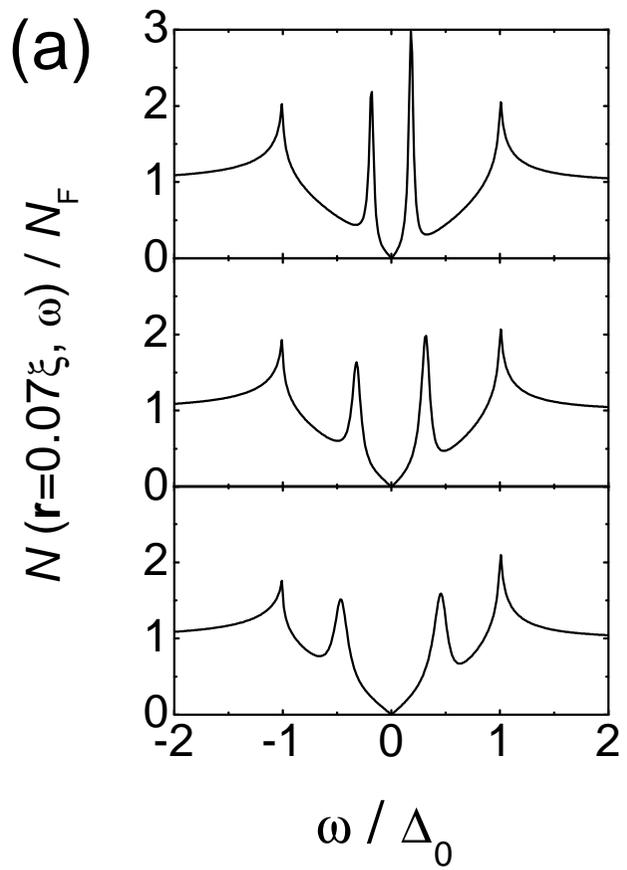
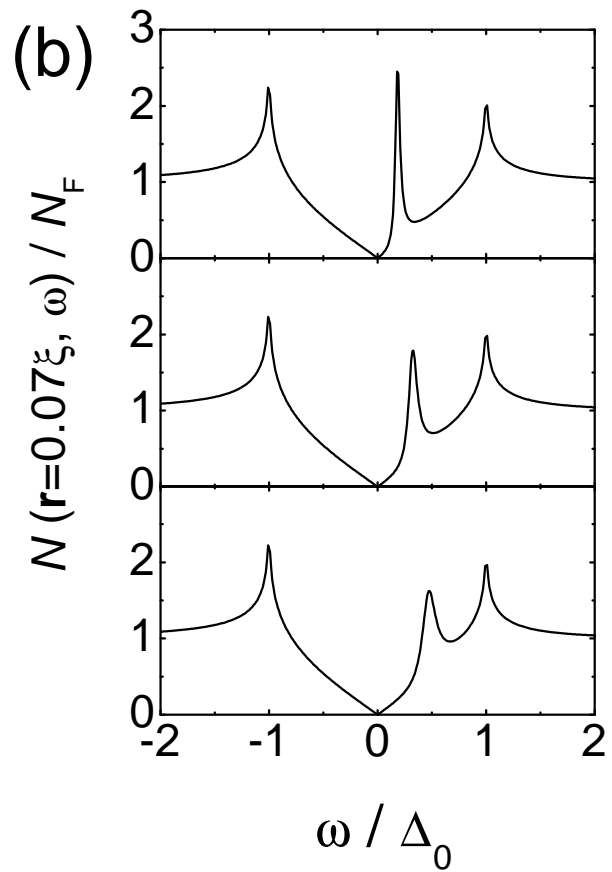
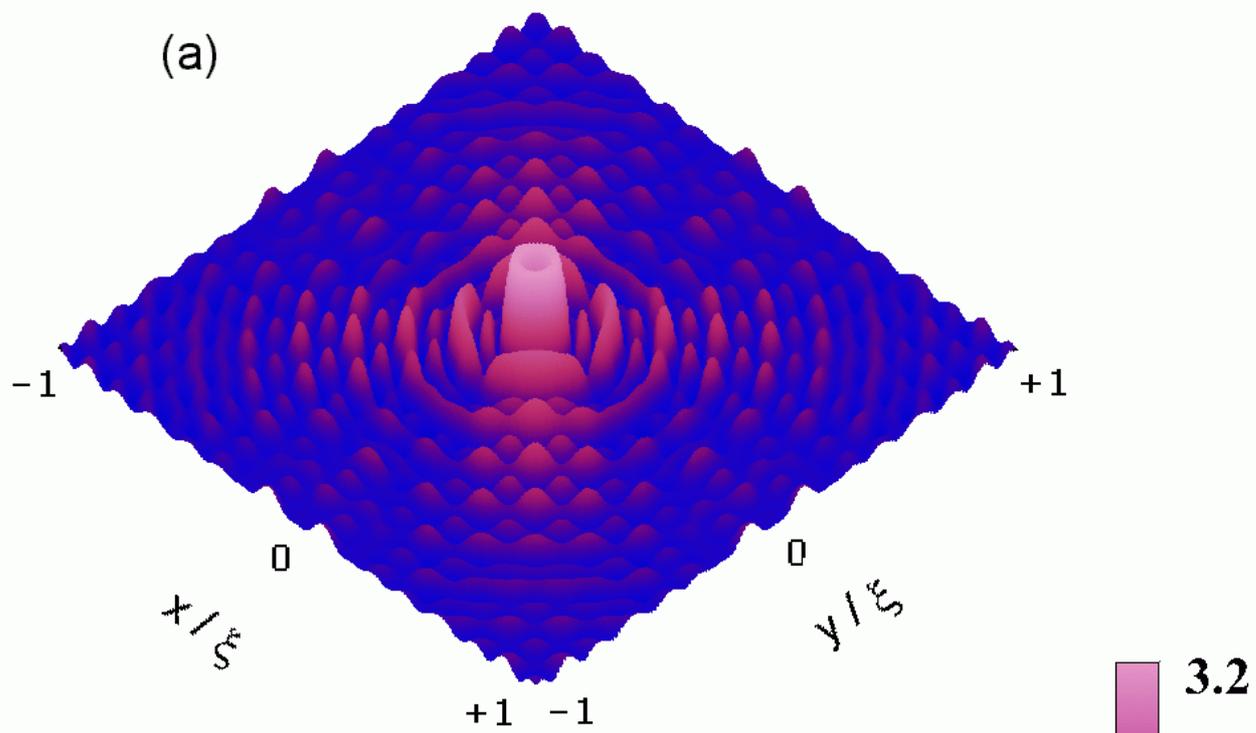
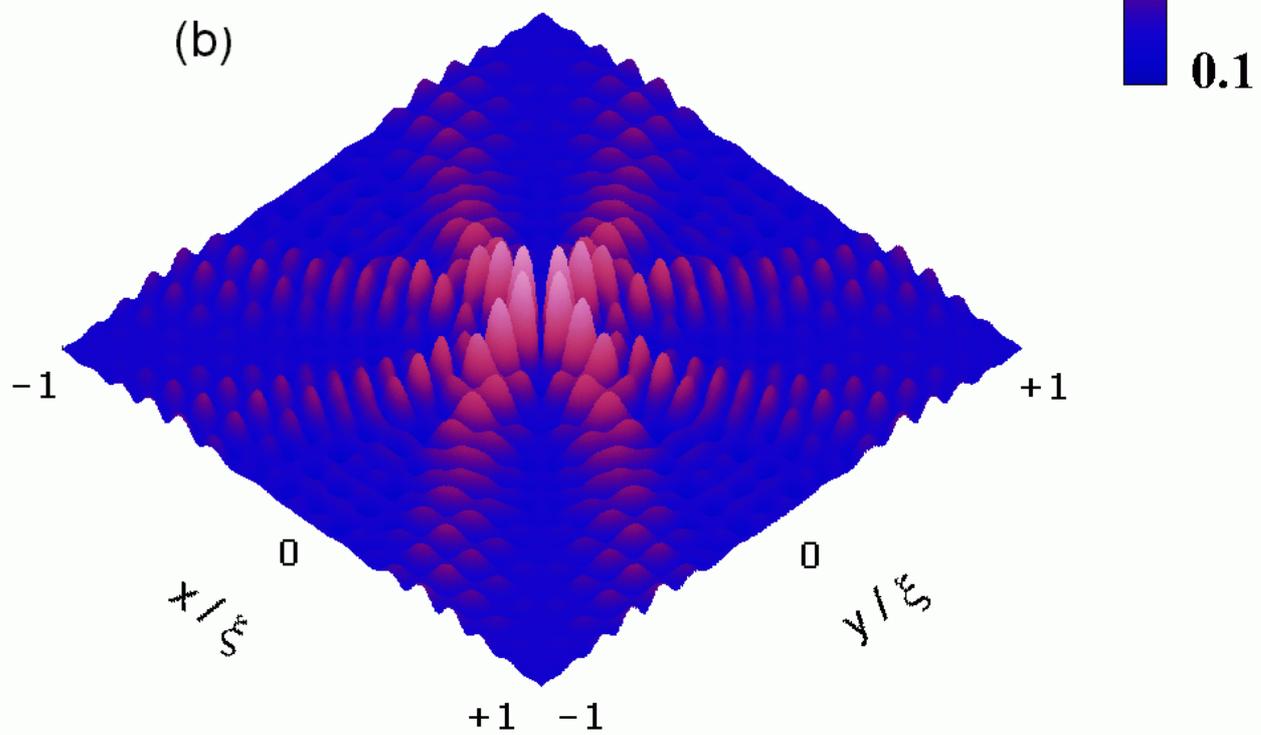
\end{document}